\documentclass[12pt,preprint]{aastex}
\shorttitle{}
\shortauthors{Nesvorn\'y et al.}
\begin{document}
\title{Photo-Dynamical Analysis of Three Kepler Objects of Interest with Significant Transit Timing Variations}
\author{David Nesvorn\'y$^1$, David Kipping$^2$, Dirk Terrell$^1$, Farhan Feroz$^3$}
\affil{(1) Department of Space Studies, Southwest Research Institute, Boulder, CO~80302, USA} 
\affil{(2) Harvard-Smithsonian Center for Astrophysics, Cambridge, MA~02138, USA}
\affil{(3) Cavendish Laboratory, University of Cambridge, Cambridge, CB3 0HE, UK}
\begin{abstract}
KOI-227, KOI-319 and KOI-884 are identified here as (at least) two planet systems. For KOI-319 and KOI-884, 
the observed Transit Timing Variations (TTVs) of the inner transiting planet are used to detect an outer 
non-transiting planet. The outer planet in KOI-884 is $\simeq$2.6 Jupiter masses and has the orbital period 
just narrow of the 3:1 resonance with the inner planet (orbital period ratio 2.93). The distribution of 
parameters inferred from KOI-319.01's TTVs is bimodal with either a $\simeq$1.6 Neptune-mass ($M_{\rm N}$) 
planet wide of the 5:3 resonance (period 80.1~d) or a $\simeq$1 Saturn-mass planet wide of the 7:3 resonance 
(period 109.2~d). The radial velocity measurements can be used in this case to determine
which of these parameter modes is correct. KOI-227.01's TTVs with large $\simeq$10 hour amplitude 
can be obtained for planetary-mass companions in various major resonances. Based on the Bayesian evidence, 
the current TTV data favor the outer 2:1 resonance with a companion mass $\simeq$1.5 $M_{\rm N}$, but this 
solution implies a very large density of KOI-227.01. The inner and outer 3:2 resonance solutions with 
sub-Neptune-mass companions are physically more plausible, but will need to be verified.    
\end{abstract}
\section{Introduction}
The Transit Timing Variations (TTVs) are deviations of transit times from a linear ephemeris that would 
be expected for a planet on a strictly Keplerian orbit. These variations can occur, for example, as a 
result of gravitational interactions of the transiting planet with a lunar, planetary or stellar companion 
(Miralda-Escud\'e 2002, Agol et al. 2005, Holman \& Murray 2005, Kipping 2008). The TTVs are important 
because they can be used to: (1) confirm and characterize planets in  
multi-transiting systems (e.g., Holman et al. 2010, Carter et al. 2012), (2) detect and characterize 
non-transiting planets in the systems where at least one planet is transiting (Nesvorn\'y et al. 2012), 
and (3) detect exomoons, some of which may be located in the habitable zone (Kipping et al. 2009). 
Item (1) is crucial for the Kepler systems with faint host stars for which the radial velocities (RVs) 
are difficult to obtain. Item (2), on the other hand, helps us to determine the incompleteness of 
the planetary systems detected from transits and understand the general distribution of mutual 
inclinations.

The detection of non-transiting planet KOI-142c (Kepler-88c) from its gravitational perturbations 
of the transiting planet KOI-142b (Kepler-88b) nicely illustrates the scientific value of TTVs (detection 
described in Nesvorn\'y et al. (2013) and KOI-142c recently confirmed from RV measurements in Barros 
et al. (2014)). In this case, the TTVs of KOI-142b were used to compute the mass of KOI-142c with a 5\% 
precision and that of KOI-142b with a 25\% precision. Combined with the physical radius of KOI-142b obtained 
from transits, these results imply the bulk density of KOI-142b to be similar to that of Saturn. The 
eccentricities and mutual inclination of the two orbits inferred from TTVs are small and very 
well determined as well. The orbital period ratio $P_c/P_b=2.03$ places the two planets just wide of the 
2:1 resonance. The properties of the KOI-142 system inferred from TTVs can therefore be useful 
to test various formation theories that have been proposed to explain the near-resonant pairs of exoplanets.

Here we continue reporting the results of our photo-dynamical analysis program to confirm and characterize 
Kepler Objects of Interests (KOIs; Borucki et al. 2011, Batalha et al. 2013) with significant TTVs. This 
is part of a broader effort known as the Hunt for Exomoons with Kepler (HEK) project (Kipping et al. 2012). 
Previous results of the project were reported in Kipping et al. (2013a,b; 2014) and Nesvorn\'y et al. (2012, 
2013). Here we discuss additional KOIs, namely KOI-227, KOI-319 and KOI-884, for which our analysis indicates 
a unique or near-unique detection of a previously unknown non-transiting object. In each case, the masses 
of the transiting and non-transiting objects obtained from the TTVs and stability analysis are firmly in 
the planetary range. KOI-227, KOI-319 and KOI-884 are therefore identified here as systems of (at least) 
two planets. Sections 2 and 3 explain our methods and results, respectively. We determine the orbital 
parameters of planets in each system and discuss how the success of the TTV-inversion method depends on 
the measured TTV properties.
\section{Methods}
A detailed description of the methods employed here can be found in Kipping et al. (2012, 2013a,b) and 
Nesvorn\'y (2012, 2013). Briefly, our analysis consists of two steps. In the first step, we fit a transit model 
to the Kepler photometry. This gives us time, duration, and other basic parameters of each transit, and their 
relative uncertainties. In the second step, we attempt to model the inferred transit times and durations 
as being caused by the gravitational interaction with a planetary, stellar or lunar companion.\footnote{The moon 
fits did not provide any plausible solutions to the TTVs of the studied systems and were ruled out based on 
the Bayesian analysis. We therefore do not discuss these fits in detail.} Note that we 
do not require that the companion itself to be transiting, although cases with multiple transiting bodies 
can be accommodated as well (e.g., Kipping et al. 2014). This two-step method is a useful alternative to a 
general photo-dynamical method, where a $N$-body integration scheme is used to directly fit the Kepler photometry 
(e.g., Carter et al. 2012). The main advantage of the general method is that it fits for a smaller number of 
parameters and more directly links computed uncertainties to the photometry. The main advantage of the 
two-step method is its transparency, which makes it easier to understand of how different dynamical 
parameters relate to the measured properties of transits. 
%
\subsection{Step 1 - Photometry}
We downloaded the publicly available Simple Aperture Photometry (SAP) data from quarters 0-17 from the Mikulski 
Archive for Space Telescopes (MAST). 
The short-cadence (SC) data were used where available, and long-cadence (LC) data otherwise. We 
extracted data near each transit and applied the Cosine Filtering with Autocorrelation Minimization 
(CoFiAM; Kipping et al. 2013a) algorithm to remove long-term trends due to effects such as focus-drift 
and rotational modulations. During this process, any other transiting signals were removed, as well as
any pointing tweaks, flares, safe-mode exponential ramps or other discontinuous features not associated
with the transit events of interest. 

After detrending the data, the next step was to fit a transit model 
to the cleaned, normalized photometry. For this purpose, we used the Mandel \& Agol (2002) algorithm 
to model the transits and the multimodal nested sampling algorithm {\tt MultiNest} (Feroz \& Hobson 2007;
Feroz et al. 2009, 2013) for the regression. Transits were fitted with 4000 live points with a target 
efficiency of 0.1. In all cases, a quadratic-law intensity profile was assumed for the limb darkening 
of the host star. Since transits are fitted on an individual basis, the limb darkening coefficients are 
generally non-convergent if freely fitted and thus we opt to fix them to values interpolated from Claret 
\& Bloemen (2011) PHOENIX models (assumed limb darkening coefficients are provided in Table~\ref{params}).
Blending from nearby sources was accounted for using the quarter-to-quarter contamination factors available 
on MAST. 

Transits were fitted with four parameters: the radius ratio $R_p/R_*$, the impact parameter $b$, the stellar density 
$\rho_*$ and the mid-transit time $\tau$. Each parameter was fitted with a uniform prior, except $\rho_*$ for which we 
adopted Jeffrey's prior between 1 and $10^6$ kg m$^{-3}$. An averaged $R_p/R_*$, $b$ and $\rho_*$ value is provided 
in Table \ref{params} for each KOI. This value was derived from an averaged posterior distribution constructed from the $N$ 
transits available. This was done by stepping through each posterior sample entry and taking the median across all 
$N$ transits. We also obtained duration of each transit, where duration is defined as the time it takes for 
the planet's center to cross the stellar disk ($\tilde{T}$; Kipping et al. 2010). The transit times and durations
of KOI-227.01, KOI-319.01 and KOI-884.02 are reported in Table \ref{td}. 

The TTVs ($\delta \tau_i$) and Transit Duration Variations (TDVs; $\delta \tilde{T}_i$) were obtained, 
respectively, by subtracting a linear ephemeris from the computed transit times, and by subtracting the mean from 
transit durations.\footnote{For KOI-227, one transit epoch was removed as an outlier since 
it had a $\tau$ uncertainty more than five times greater than the median $\tau$ uncertainty. The same filter 
on KOI-884 removed 12 transits, but none for KOI-314.}
These planet candidates were selected from our previous analysis of 
the TTVs published by Mazeh et al. (2013) as interesting cases where the TTVs can potentially yield planet 
confirmation. Table \ref{params} shows the parameters of the selected candidates derived from our 
transit analysis.

In order to convert from relative masses and radii to physical units, we require an input stellar mass and radius.
To this end, we adopt the stellar parameters derived recently by Huber et al. (2014) and we direct those 
interested to this paper for details. The relevant stellar parameters used in this work are provided in 
Table~\ref{stellar}. 
\subsection{Step 2 - Dynamics}
Dynamical fits were obtained with a code based on a symplectic $N$-body integrator known as {\tt swift\_{}mvs} 
(Levison \& Duncan 1994), which is an efficient implementation of the second-order map developed 
by Wisdom \& Holman (1991). The integration was done in Jacobi coordinates. We used a symplectic corrector 
(Wisdom 2006). The code computes the mid-transit times by interpolation as described in Nesvorn\'y et al. 
(2013) and Deck et al. (2014). With an integration time step of 1/20 of the orbital period, the typical 
precision is better than a few seconds, which is better than needed because the TTV measurements have $>$1 
minute errors.

The code initially attempts to provide the best fit to the TTVs and TDVs determined in Step~1 (instead of considering 
the actual transit times and durations). This model, hereafter ${\cal M}_{\rm V}$, is useful for a non-transiting 
perturber where the orbital period of the perturber is not constrained from the transit ephemeris itself. In such a 
case, we find it best to remove a linear ephemeris from the transit times of the transiting planet candidate, and not 
to be concerned with it when performing a dynamical fit. This helps to avoid various potential problems, because a 
linear trend of $\delta t$ can be produced by a number of effects that are not modeled (e.g., secular perturbations 
from a 3rd object, tidal or relativistic precession). 

In the next step, the code attempts to provide a more general fit to the transit times $\tau_i$ and transit 
durations $\tilde{T}_i$. This has the advantage that the orbital period of the transiting object is constrained from 
the transit ephemeris, and the dynamical fit automatically yields a correct orbital period of the companion (and 
not only the period ratio). A slight disadvantage of the general fit is that it requires a larger number of parameters, and 
therefore takes longer to converge. Also, the general fit suppresses a clear distinction between the photometric and 
dynamical steps of our analysis, because some parameters, namely the orbital period and impact parameter of the 
transiting object, are computed in both steps. This can be used to test the consistency of results. 

In practice, we prefer to derive the final results with a hybrid model, ${\cal M}_{\rm H}$, 
in which the code fits for the transit times ($\tau_i$) 
and TDVs ($\delta \tilde{T}_i$). Model ${\cal M}_{\rm H}$ maximizes the dynamical information contained in the measured 
transit timing, including both the linear transit ephemeris and TTVs, and does not introduce additional parameters 
that would be needed for modeling of transit durations (e.g., impact parameter, stellar radius). It is especially 
useful in this study, because the selected systems do not show any evidence for TDVs, and therefore not much dynamical 
information can be extracted from the duration measurements.

${\cal M}_{\rm V}$ has ten parameters: mass ratios $M_b/M_*$ and $M_c/M_*$, orbital period ratio $P_c/P_b$, 
eccentricities $e_b$ and $e_c$, pericenter longitudes $\varpi_b$ and $\varpi_c$, mean longitude $\lambda_c$, nodal 
longitude difference $\Delta \Omega = \Omega_c-\Omega_b$, and inclination $i_c$, where indices $b$ and $c$ denote 
the transiter and perturber, respectively. Uniform priors were used for all parameters, except $i_c$ for which
we used an isotropic prior. 
${\cal M}_{\rm H}$ has twelve parameters. In addition to those included in 
${\cal M}_{\rm V}$, there is the orbital period $P_b$, and $t_b$, where $t_b$ denotes the time interval between
the reference epoch $\tau_0$ and the mid-transit time of a selected transit. Note that the parameters $P_b$ and 
$P_c$ are the {\it osculating} periods at epoch $\tau_0$, while the orbital period $P$ determined from the transit 
ephemeris in the photometric step (Section 2.1, Table \ref{params}) is the {\it mean} orbital period.  
 
We used the transit reference system (Nesvorn\'y et al. 2012), where the nodal longitude is $\Omega_b=270^\circ$ 
by definition, and $i=0$ for the impact parameter $b=0$. The time epoch $\tau_0$ was chosen very near of the 
mid-transit time of a selected transit. We examined periods between 1 day and 10 yr, including cases 
of highly eccentric and/or retrograde orbits. 

To maximize the likelihood of the fits, or equivalently to minimize $\chi^2$, we used the Downhill Simplex Method 
(DSM; Press et al. 1992) and the {\tt MultiNest} algorithm (Feroz \& Hobson 2007; Feroz et al. 2009, 2013). 
The DSM was adapted to 
this problem in Nesvorn\'y \& Beaug\'e (2010). It is expected to find the correct solution if the transit coverage 
and timing uncertainties are adequate. We use the DSM with up to $10^7$ chains to make sure that none of the potential 
solutions has been missed by {\tt MultiNest}. The {\tt MultiNest} is a multimodal nested sampling routine 
designed to compute the Bayesian evidence in complex and multimodal parameter space (including curved 
degeneracies) in efficient manner. Although computing the evidence is the primary goal of the code,
which is valuable in later model selection, it also produces the posterior parameter distribution as a by-product.
Marginalizing these posteriors allows us to compute the nominal parameter estimate (for which we use the median)
and the associated uncertainties (for which we use the 34.1\% quantiles on either side of the median).
\section{Results}
KOI-227, KOI-319 and KOI-884 are discussed below in a reverse order. We start with the best case, KOI-884, 
where the determination of parameters is unique (except for the mutual inclination of the two orbits). We then 
move to KOI-319, where the distribution of dynamical parameters inferred from the TTVs is bimodal. Finally,
we discuss KOI-227, where the existing TTVs imply a planetary companion in one of the major resonances, 
most likely 2:1 or 3:2, with the transiting planet. 
\subsection{KOI-884}
Previous works listed KOI-884 as having two or three transiting planet candidates (Ford et al. 2011, 
2012; Mazeh et al. 2013). The estimated periods of the inner orbits are 3.3 and 9.4 days. Here we 
analyzed transits of the outer candidate, KOI-884.02, which shows transits with a 20.48-day 
period (Table \ref{params}). 

\begin{figure}[t]
\epsscale{0.6}
\plotone{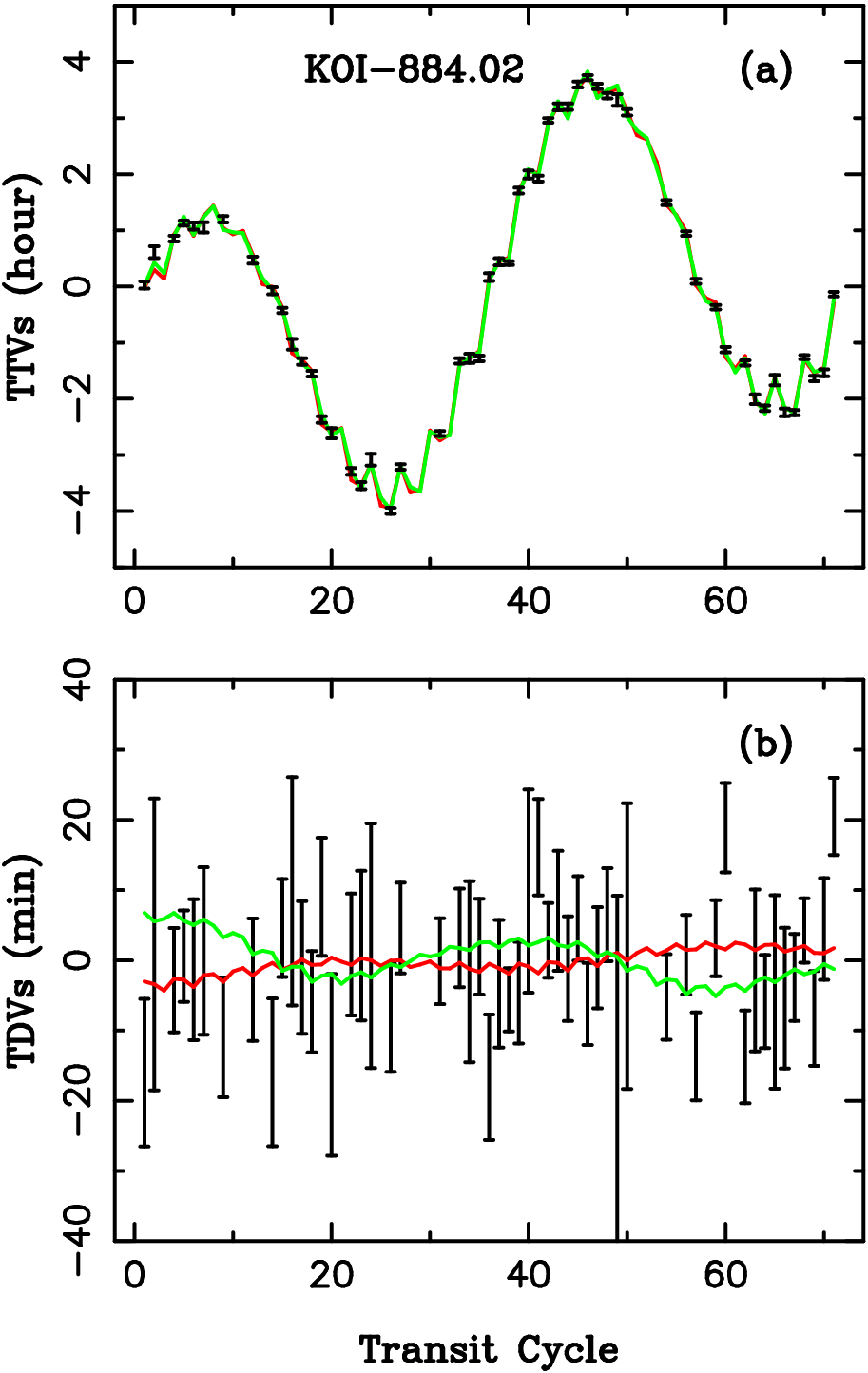}
\caption{TTVs and TDVs of KOI-884.02. The error bars show the measured TTVs and TDVs
and their uncertainty. The red and green lines show the best fits corresponding to
Modes 1 and 2, respectively. The lines nearly overlap in panel (a).}
\label{koi884}
\end{figure}

Our photometric analysis yielded 54 usable transits of KOI-884.02, for which accurate transit parameters have 
been determined. Large, nearly-sinusoidal TTVs are detected with a nearly 4 hour amplitude and 
$\simeq$830 day period (Figure \ref{koi884}a). The TTV signal contains short-period `chopping', which 
can be indicative of conjunctions. The chopping is the most obvious for transit cycles between 32 to 45, where 
three subsequent transits on a nearly linear ephemeris are offset from the next 
three transits by up to $\simeq$100 min. We found no evidence for the TDVs (Figure \ref{koi884}b).

To start with, we restricted the dynamical analysis of KOI-884.02's TTVs to the companion periods 
indicated by the inner planet candidates (i.e., near 3.3 or 9.4 days). We found no plausible solutions 
in this case, leading us to conclude that the observed TTVs of KOI-884.02 cannot be explained by 
KOI-884.01 nor KOI-884.03.

Our global fit was performed over the full range of companion periods. It showed that the observed variations can 
be explained by a companion on an outer orbit near the 3:1 resonance (orbital period $\simeq$59.96 day). When 
model ${\cal M}_{\rm V}$ was used to constrain the companion properties, the fit produced $\chi^2=213.5$ for 
$108-10=98$ degrees of freedom (DOF). With ${\cal M}_{\rm H}$ we obtained $\chi^2=213.3$ for $108-12=96$. 
These results are satisfactory (Figure \ref{koi884}) and may suggest either slightly underestimated timing 
uncertainties or additional noise due to stellar activity or additional low-amplitude perturbations. Similarly, 
a TTV-only fit gives $\chi^2=80.9$ for 44 DOF. The results of the TTV-only fit, however, imply the high mutual 
inclination of orbits ($i_{\rm Mutual}\simeq14\degr$) and strong TDVs that are not observed. Below we discuss the 
results obtained from ${\cal M}_{\rm H}$.

During an earlier analysis of the timing data from quarters 0-12, our modeling suggested the possibility that 
the observed variations can be explained by a companion near the exterior 3:2 resonance (orbital period $\simeq$31.1 d). 
Including quarters up to 17, however, these 3:2 solutions appear implausible since they yield $\chi^2>1000$ 
for 96 DOF, and $\Delta \chi^2>1000-213=787$ relative to the best-fit solution near the 3:1 resonance.
We therefore conclude that they can be ruled out at high confidence. Additionally, comparing the evidences 
obtained from the {\tt MultiNest}, we find $\Delta \ln {\cal Z}>100$, corresponding to an overwhelming statistical 
preference for the 3:1 solution.\footnote{Companions near other resonances, such as the 2:1, 4:1 or 5:1 outer MMRs, 
can fit the sinusoidal TTV oscillation reasonably well, but these resonances fail at producing the observed 
short-period chopping. This explains why these resonances yielded unsatisfactory results.} Other parameter modes, 
including companions on interior, highly-inclined or retrograde orbits, were also ruled out, because they these 
solutions have $\chi^2>1000$ and are disfavored relative to the 3:1 solution by $\Delta \ln {\cal Z}>100$. 
     
We therefore find that the dynamical analysis of KOI-884.02's transits allows us to uniquely infer an outer 
companion near the 3:1 resonance. The estimated mass of the companion is $2.67_{-0.26}^{+0.38}\times 10^{-3}\ M_\star$.  
With $M_\star=0.884\ M_\odot$ from Huber et al. (2014), this corresponds to $\simeq$2.4 Jupiter masses 
($M_{\rm J}$). The mass of the transiting object is also constrained from ${\cal M}_{\rm H}$ 
to be $<$1.8 $M_{\rm J}$ (99\% confidence). 
In addition, our stability analysis indicates that the system would not be stable for masses $>$5-17 $M_{\rm J}$
(the exact value of this cutoff depends on the orbital eccentricities). Together, these results demonstrate 
that KOI-884 is a system of (at least) two planets, hereafter KOI-884b (corresponding to KOI-884.02) and 
non-transiting KOI-884c.

\begin{figure}[t]
\epsscale{1.0}
\plotone{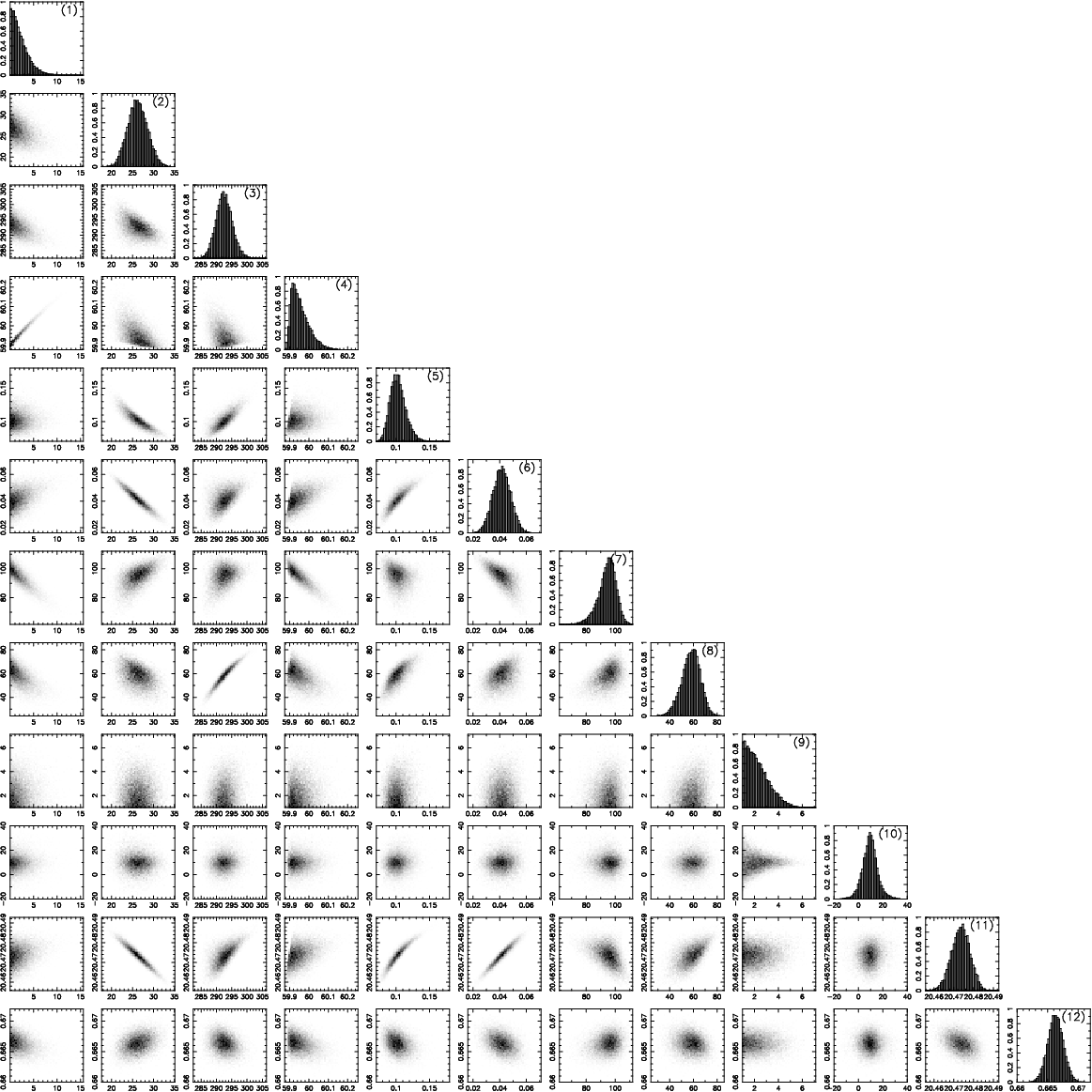}
\caption{The posterior distribution of dynamical parameters for Mode 1 of KOI-884. See Table
\ref{koi884_tab} for the definition of parameters shown here. The masses in panels (1) and (2) were scaled 
by a factor of $10^{-4}$.}
\label{koi884_lowi}
\end{figure}

\begin{figure}[t]
\epsscale{1.0}
\plotone{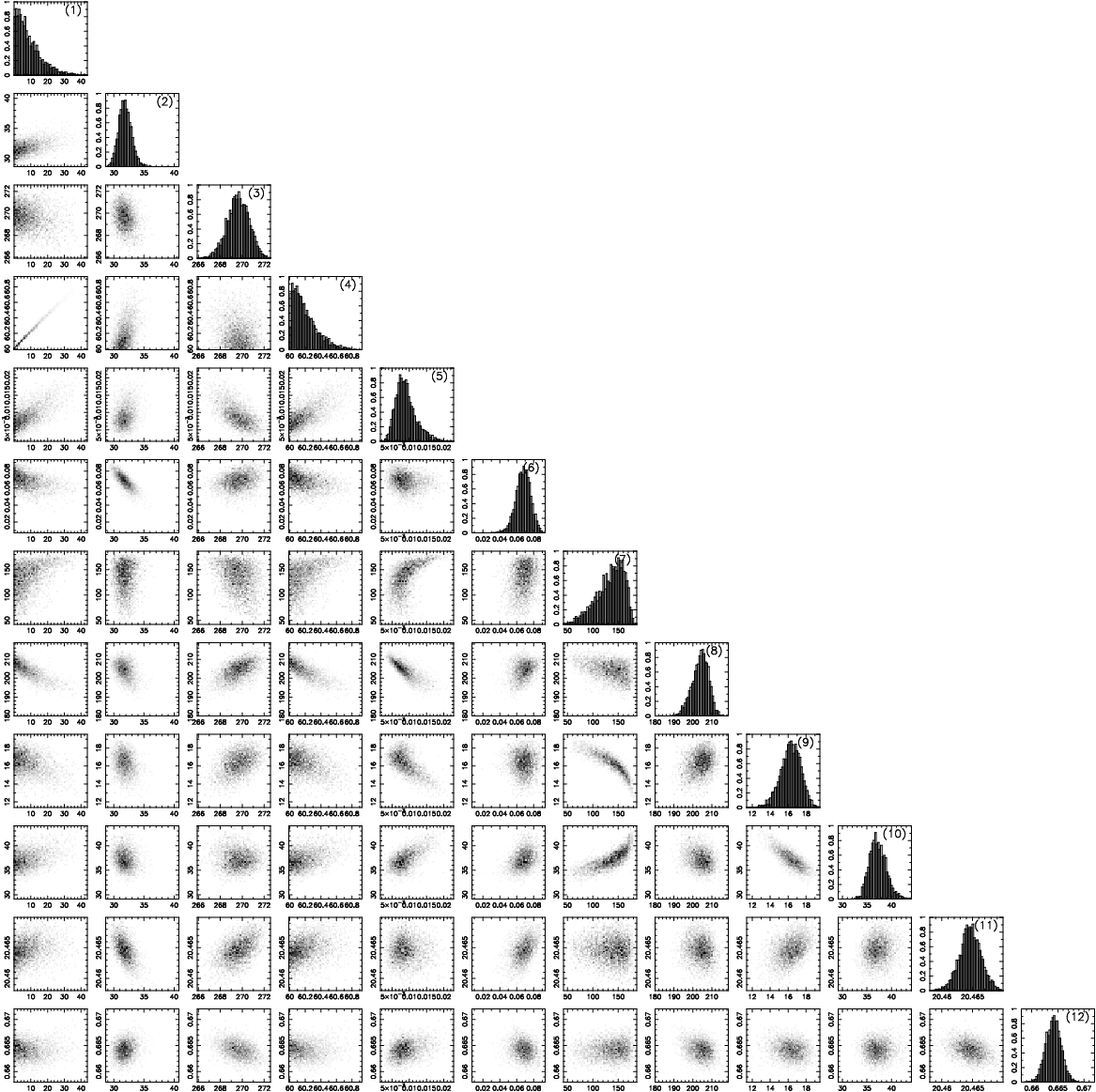}
\caption{The posterior distribution of dynamical parameters for Mode 2 of KOI-884. See Table
\ref{koi884_tab} for the definition of parameters shown here. The masses  in panels (1) and (2) were scaled 
by a factor of $10^{-4}$.}
\label{koi884_higi}
\end{figure}

The posterior distribution of parameters obtained from {\tt MultiNest} reveals the existence of
two modes: Mode 1 with low impact parameter values of the outer companion ($b_c=2.11_{-0.79}^{+1.18}$),
and Mode 2 with high values ($b_c=16.3_{-1.2}^{+1.1}$), implying larger mutual inclination
between the two orbits. The evidence terms for the two modes are $\ln {\cal Z}=349.2$ and
$\ln {\cal Z}=347.4$, respectively. This shows a slight preference for Mode 1, but the difference
is clearly inconclusive. Below we discuss the two modes separately. Figures \ref{koi884_lowi} and 
\ref{koi884_higi} show the posterior distribution of dynamical parameters for Modes 1 and 2,
and Table \ref{koi884_tab} provides the best fit parameter values and their uncertainties. 

The outer planet mass corresponding to Mode 1 is somewhat lower ($M_c\simeq2.4$ $M_{\rm J}$) 
than the one corresponding to Mode 2 ($M_c\simeq2.9$ $M_{\rm J}$). As for the transiting planet mass, 
$M_b=50_{-35}^{+65}\, M_\oplus$ for Mode 1 (essentially an upper limit; see Figure \ref{koi884_lowi}). 
Mode 2, instead, implies that $M_b = 200_{-140}^{+270}\, M_\oplus$. We point out that the lower mass values of 
Mode 1 appear to better relate to $R_b\simeq4.5\,R_\oplus$ inferred from the transit depth, giving 
$\rho_b=3.1_{-2.2}^{+4.0}$ g cm$^{-3}$. Mode 2, on the other hand, would require a very high density 
of KOI-884b, $\rho_b=12.2_{-4.2}^{+6.6}$ g cm$^{-3}$, which would clearly be too high for a 
Saturn-class planet. 
  
Mode 1 implies a very low mutual inclination ($i_{\rm Mutual}\simeq0.75^\circ$). The orbits cannot be 
precisely coplanar because the transits of KOI-884c are not detected. A small mutual inclination, 
$i_{\rm Mutual}\simeq0.2\degr$, however, would be sufficient for KOI-884c to avoid transits. The mutual 
inclination is much larger in Mode~2 ($i_{\rm Mutual}\simeq13.5^\circ$). According to our tests, such a 
substantial inclination would normally lead to important TDVs, but Mode 2 also fixes $\Delta \Omega$ 
such that the immediate effects of the nodal precession on $b_b$ are limited. Mode 2 therefore 
does not lead to significant TDVs, as shown in Figure \ref{koi884}b. 

According to Figueira et al. (2012) the mutual inclination of exoplanets follows a Rayleigh distribution 
with mode at 1$^\circ$ nominally (and $5^\circ$ conservatively; see also Fang \& Margot 2012). This means that 
$i_{\rm Mutual}>10^\circ$ would be almost 10-sigma less likely than $i_{\rm Mutual}<10^\circ$. This 
argument, and the one discussed above for $\rho_b$, could be used to favor Mode 1 {\it a priori},
but we do not draw any firm conclusion from this argument here.   
  
The orbital eccentricities of both planets are relatively small. Mode 1 gives a larger 
eccentricity value for KOI-884b at the reference epoch ($e_b\simeq0.1$) than Mode~2 ($e_b\simeq0.009$). 
In Mode 1, $e_b$ has secular oscillations with a $\simeq$90~yr period, large amplitude, 
and mean value of $\simeq$0.07. In Mode 2, the oscillations of $e_b$ are faster and smaller. 
The variations of $e_c$ are much smaller, because of the large mass of KOI-884c.  

The two planets are apparently very close to but not {\it inside} 
the 3:1 resonance.\footnote{A small change of orbital elements would bring the two planets inside the resonance.
This would be accomplished, for example, by decreasing the semimajor axis $a_b$ of Mode 1 by only $\simeq$1.5\%.}
The orbital period ratio suggested by both modes is $P_c/P_b=2.93$, slightly smaller 
than that of the exact resonance. The the 3:1 resonance angle, $3\lambda_c-1\lambda_b$, circulates 
in the prograde sense with period $1/(3/P_c-1/P_b)\simeq830$ days, which is the dominant period detected 
in the TTVs. 
\subsection{KOI-319}
\begin{figure}[t]
\epsscale{0.6}
\plotone{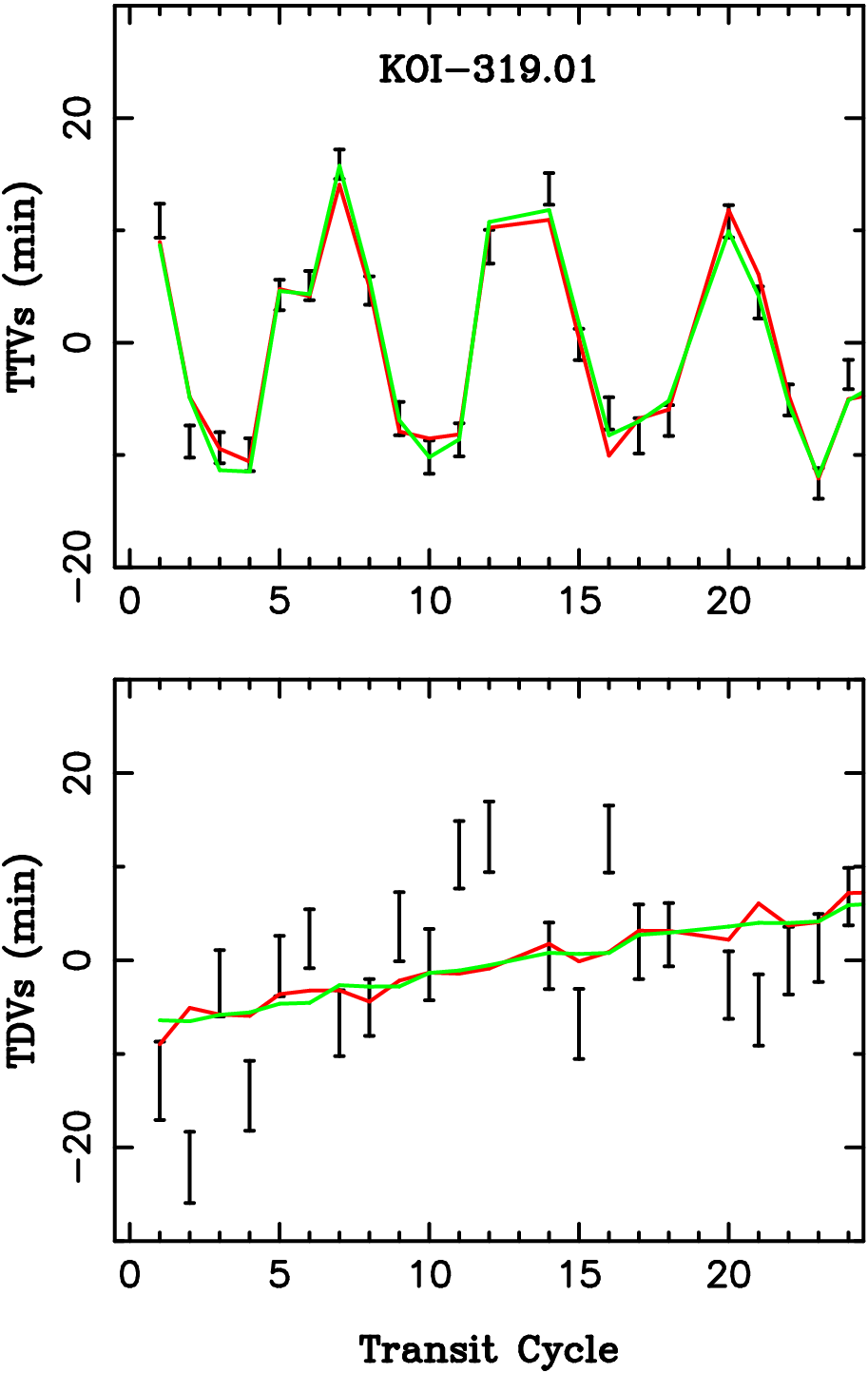}
\caption{TTVs and TDVs of KOI-319.01. The error bars show the measured TTVs and TDVs
and their uncertainty. The red and green lines show the best fits corresponding to
Modes 1 and 2, respectively.}
\label{koi319_fit}
\end{figure}

KOI-319.01's transits occur with the period of $\simeq$46.15 days. The measured TTVs have a relatively small 
amplitude ($\simeq$12 min) and apparently contain more than one frequency (Figure~\ref{koi319_fit}), which may
be a signature of short-periodic variations caused by a companion. 

Our dynamical analysis with the DSM identifies two planetary solutions that fit data (nearly) equally well. 
Mode 1 (hereafter M1) has $\chi^2=116.5$ and mode 2 (M2) has $\chi^2=126.3$, both for 36 DOF in 
${\cal M}_{\rm H}$. These $\chi^2$ 
values are larger than ideal, but this can be understood because the fluctuating pattern of TDVs in 
Figure~\ref{koi319_fit}b is difficult to fit by dynamics. Instead, our TTV-only fits indicate that TDVs should 
either be absent or having a slow linear trend. Therefore, either the TDV fluctuations are not real (implying 
that the TDV errors were slightly underestimated) or are caused by some other effect. The TTV-only fits 
give $\chi^2=28.5$ for M1 and $\chi^2=29.4$ for M2, both for 14 DOF. Other TTV-only fits have $\chi^2>50$ 
and can formally be excluded with $>$90\% confidence, which is suggestive (but not 
conclusive).\footnote{For example, the best inner solutions with $\simeq17.9$ and $\simeq19.1$ periods have 
$\chi^2=53.0$ and $\chi^2=59.4$.}
 
In a global dynamical fit, where the orbital period priors were chosen to range between 1 day and 10 years, 
the {\tt MultiNest} generates a unimodal posterior distribution of periods corresponding to that of M2. 
To identify M1, the orbital period range needs to be restricted by the selection of priors around periods 
corresponding to M1, and excluding M2. This happens because the likelihood landscape favors detection
of M2, and the {\tt MultiNest} code (we use 4000 or 8000 live points) converges to M2 if the period prior includes 
M2. The M1 mode is slightly preferred using the Bayesian evidence ($\ln {\cal Z}=171.8$, where ${\cal Z}$
is the evidence) over the M2 mode ($\ln {\cal Z}=169.8$), but this difference is clearly inconclusive. 
We discuss both M1 and M2 below.\footnote{The moon fits to the TTVs of KOI-319b did not provide any 
plausible results. With $\ln {\cal Z}=105.8$, they can be ruled out based on the overwhelming statistical 
preference for M1 and M2.}

Our dynamical analysis with ${\cal M}_{\rm H}$ implies that the companion mass is $\simeq 6.3\times10^{-5}\ M_\star$ for M1 
and $2.4\times10^{-4}\ M_\star$ for M2. With $M_\star \simeq 1.3\ M_\odot$ from Huber et al. (2014), this
yields masses $\simeq1.6\ M_{\rm N}$ and $\simeq1.1\, M_{\rm S}$, respectively, where $M_{\rm S}$ and
$M_{\rm N}$ are the masses of Saturn and Neptune. The mass of the transiting object 
cannot be constrained from the existing TTVs (or TDVs) alone, but our stability analysis implies that it is 
$<10\, M_{\rm J}$ for M1 and $<16\, M_{\rm J}$ for M2. We therefore find that KOI-319 is a system of (at least) 
two planets, hereafter KOI-319b, corresponding to KOI-319.01, and KOI-319c on an outer orbit.

\begin{figure}[t]
\epsscale{1.0}
\plotone{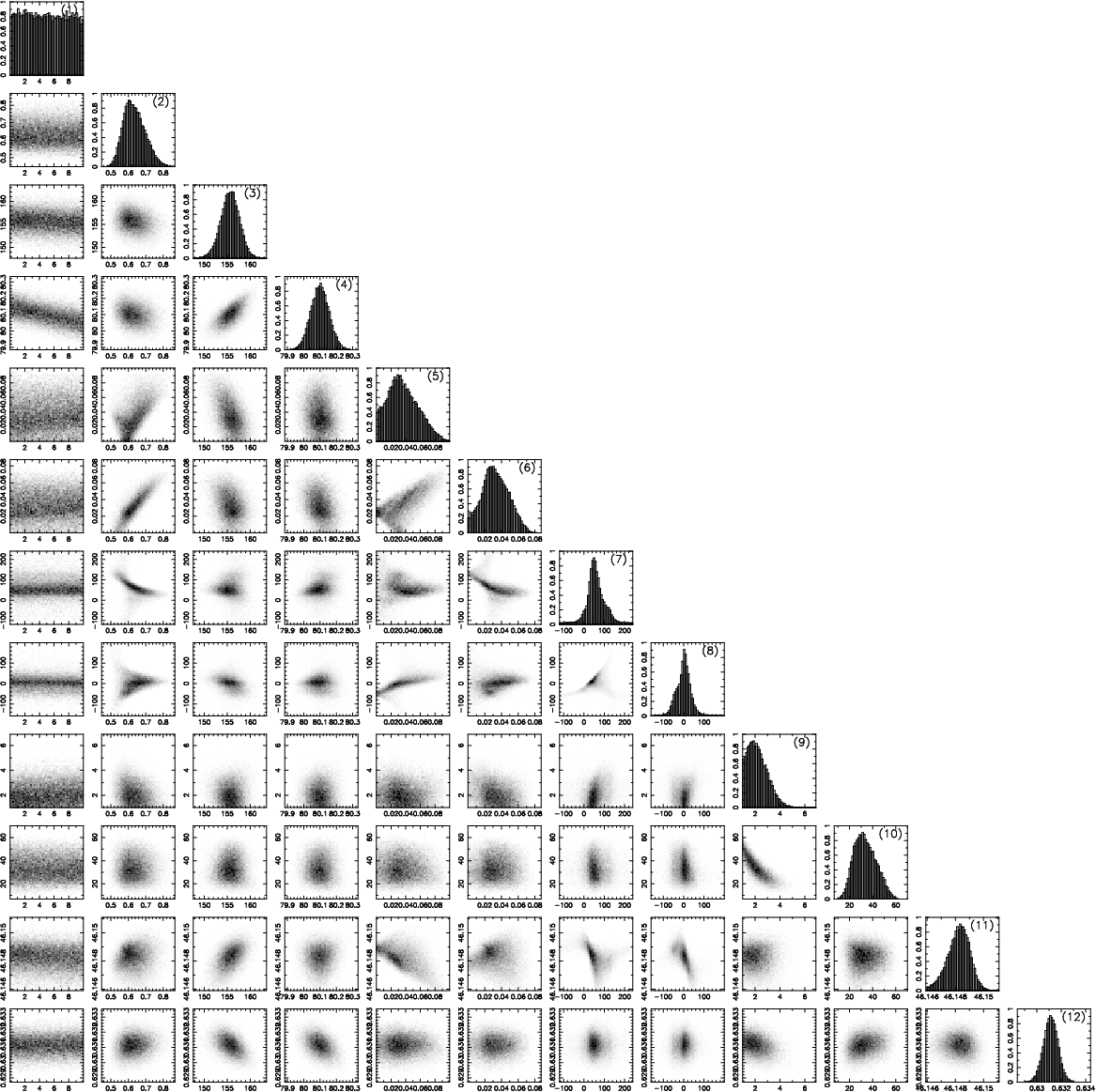}
\caption{The posterior distribution of dynamical parameters for Mode 1 of KOI-319. See Table
\ref{koi319_tab} for the definition of parameters shown here. The masses in panels (1) and (2) were scaled 
by a factor of $10^{-4}$.}
\label{koi319_mode1}
\end{figure}

\begin{figure}[t]
\epsscale{1.0}
\plotone{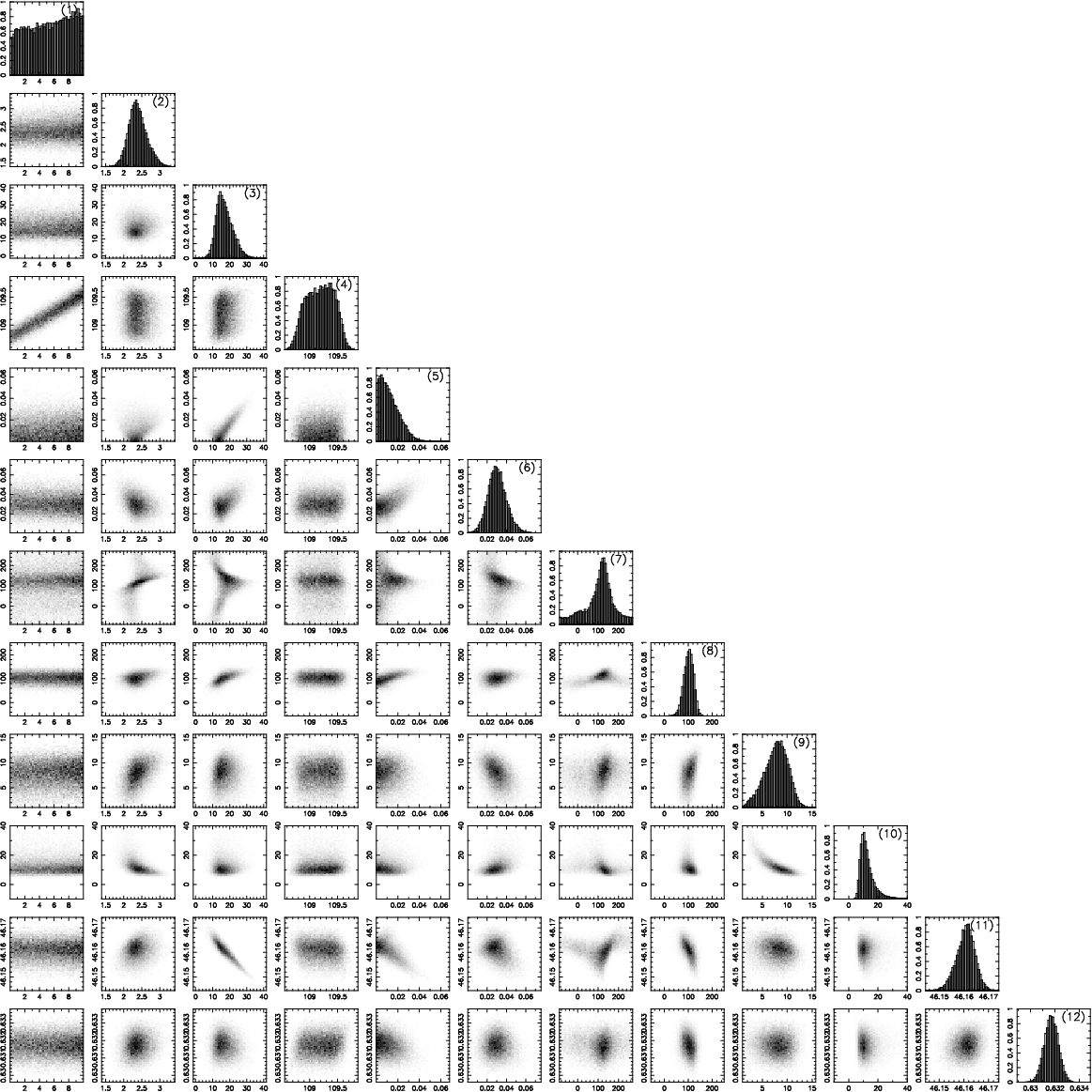}
\caption{The posterior distribution of dynamical parameters for Mode 2 of KOI-319. See Table
\ref{koi319_tab} for the definition of parameters shown here. The masses in panels (1) and (2) were scaled 
by a factor of $10^{-4}$.}
\label{koi319_mode2}
\end{figure}

The posterior distributions of the two modes are shown in Figures \ref{koi319_mode1} and \ref{koi319_mode2}. 
The best fit parameters and their uncertainties from ${\cal M}_{\rm H}$ are reported in Table \ref{koi319_tab}. 
M2 gives $M_c\simeq104\, M_\oplus$ and $P_c \simeq 109.2$ day. This would imply $P_c/P_b\simeq 2.36$ with the two 
orbits just wide of the 7:3 resonance. M1 gives $M_c\simeq28\, M_\oplus$, $P_c\simeq 80.1$ day, and $P_c/P_b\simeq 
1.74$, just wide of the 5:3 resonance. M1 could be favored over M2 {\it a priori}, because the $\sim30\, M_\oplus$ 
planets are presumably more common than the $\sim100\, M_\oplus$ planets (Mayor et al. 2011), but we do not
draw any firm conclusion from this argument.

The orbital eccentricities of KOI-319b and KOI-319c indicated by the two 
solutions are likely to be small ($e<0.1$ with a 99.7\% confidence). The mutual inclination of orbits must 
be significant such that the transits of KOI-319c are not seen, but not too large, as inferred from 
from the TTV fit (e.g., $i_{\rm Mutual}<5.4^\circ$ for M1 with a 99.7\% confidence).    

The symmetric placement of the two modes identified here around the 2:1 resonance implies 
that in each case $2\lambda_c-\lambda_b$ circulates (progradely for M1 and retrogradely for 
M2) with a $\simeq300$ day period. This is the dominant TTV period seen in Figure \ref{koi319_fit}. 
More generally, the symmetric placement assures that terms of M1 with $k\lambda_b-j\lambda_c$ 
have the same frequencies as the terms of M2 with $k\lambda_b-(k-j)\lambda_c$, where $k$ 
and $j$ are integers. The amplitude of these terms is a complex function of mass, eccentricity 
and inclination of the perturber, and can apparently be adjusted in each case to fit the existing 
data very well. 

More TTV data or radial velocity (RV) measurements will be needed to validate our
results and decide which of the two parameter modes identified here is correct. 
The RV measurements should be feasible in this case, because the host star is relatively bright (12.7
Kepler magnitude). The RV term corresponding to KOI-319b should have the 46.15 day period and $\sim$50 m s$^{-1}$
amplitude (assuming that KOI-319b is a Jupiter-class planet as indicated by its large physical 
radius, $R \simeq 10.7\,R_\oplus$, inferred from the photometric analysis). The second term present 
in the RV signal should either have the 80.1 day period and $\simeq$3.5~m~s$^{-1}$ amplitude,
corresponding to M1, or the 109.2 day period and $\simeq$12 m s$^{-1}$ amplitude, corresponding 
to M2.
\subsection{KOI-227}
\begin{figure}[t]
\epsscale{0.6}
\plotone{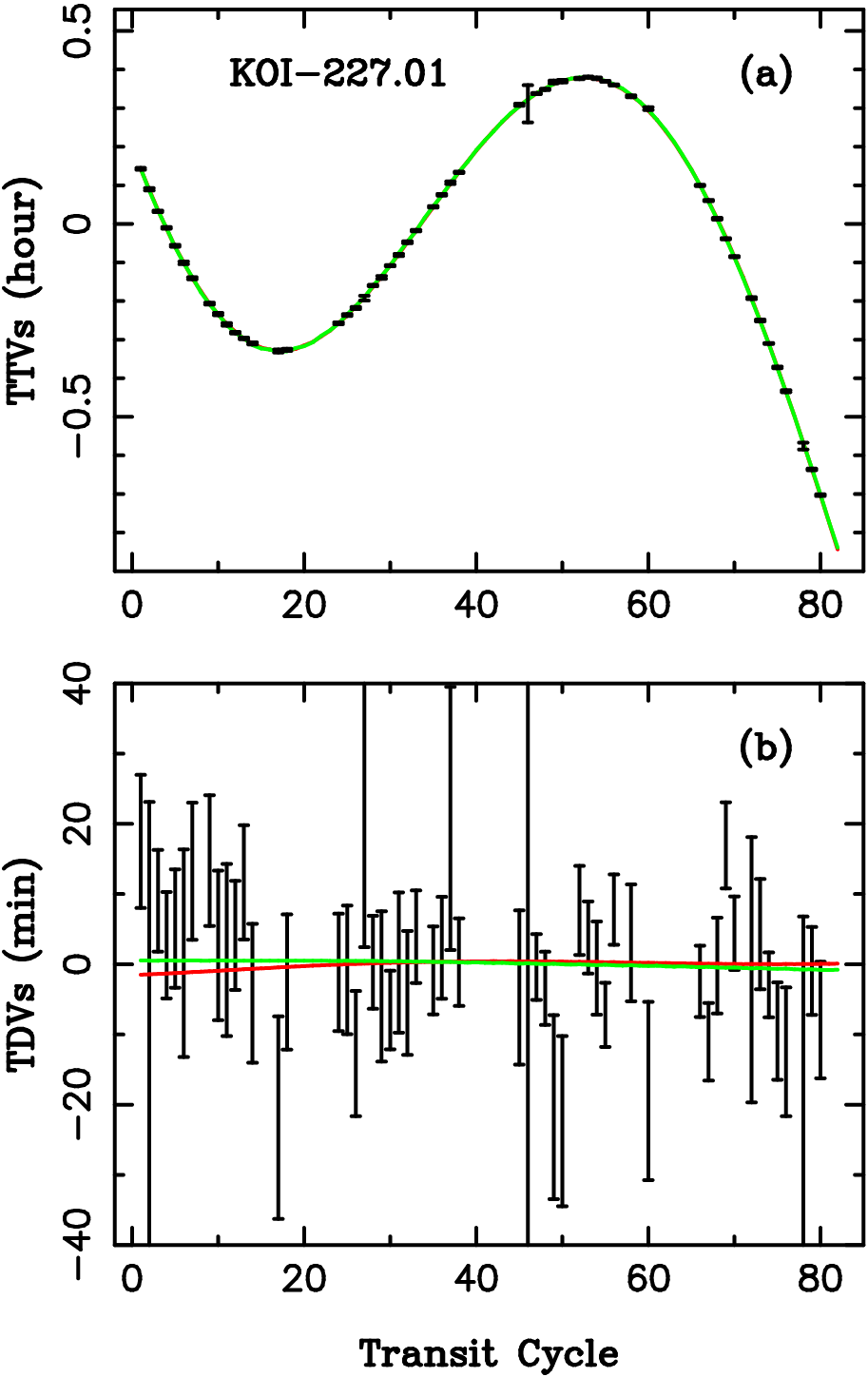}
\caption{TTVs and TDVs of KOI-227.01. The error bars show the measured TTVs and TDVs
and their uncertainty. The red and green lines show the best fits corresponding to
the 2:1 and 3:2 resonances, respectively.}
\label{koi227_fit}
\end{figure}

KOI-227.01 is a single transiter with an approximate period of 17.7 days. It shows a very large amplitude 
($\simeq10$ hours) and long period ($>1000$ days) TTVs (Figure \ref{koi227_fit}), which rival in amplitude 
those of the KOI-142b (Nesvorn\'y et al. 2013).\footnote{If KOI-142 is the ``king of TTVs'' (Mazeh et al. 
2013, Nesvorn\'y et al. 2013, Barros et al. 2014) then KOI-227 should be the ``queen''.} Such large TTVs 
may indicate the presence of a perturbing companion very close to or in a nearby major resonance with 
the transiting object.  

When the {\tt MultiNest} fit is restricted to the companion periods between 1 to 40 days, corresponding 
to all plausible inner orbits and outer orbits up to and including the 2:1 resonance, the outer 2:1 and 
3:2 resonances appear as favored solutions based on the Bayesian evidence. For example, $\ln {\cal Z}=227.7$ 
for the 2:1 mode, $\ln {\cal Z}=225.8$ for the 3:2 mode, and $\ln {\cal Z}=214.4$ for the mode 
corresponding to the outer 5:3 resonance. The two former resonances are therefore favored by healthy 
margins of $\Delta \ln {\cal Z}=13.3$ and 11.4, respectively, over the 5:3 resonance.\footnote{For a reference, 
the best TTV-only fits in the 2:1, 3:2 and 5:3 resonances have $\chi^2=37.6$, 51.5 and 82.2, respectively, 
for 45 DOF.} The solution in the inner 3:2 resonance stands out from all inner solutions with $\ln {\cal Z}=221.4$. 
Other parameter modes corresponding to orbital periods between 1 day and 40 days have $\ln {\cal Z}<210$, 
and are therefore strongly disfavored.

Our global dynamical fits for periods exceeding 40 days show additional solutions that cannot be ruled 
out based on the evidence term (they give $\ln {\cal Z}=215$-225). These parameter modes correspond to 
outer resonances such as the 3:1, 4:1 and 5:1. Interestingly, the masses required by these 
outer solutions are generally much larger than the ones required for the 2:1 resonance solution, and frequently 
(but not always) imply a very large density of the transiting object ($>$50 g cm$^{-3}$), which would 
be nonphysical. We therefore find that these parameter modes would need to be analyzed in detail to establish 
their plausibility. Here we discuss the properties of the modes corresponding to the outer 2:1 and 3:2 
resonances, and to the inner 3:2 resonance. These solutions fit data very well (Figures \ref{koi227_fit} 
and \ref{chop}) and appear to be physically more plausible.
 
Marginalized posteriors yield companion masses $\simeq1.6\times10^{-4}\ M_\star$ for the outer 2:1 mode, 
$\simeq7.7\times10^{-5}\ M_\star$ for the outer 3:2 mode, and $4.82_{-0.84}^{+1.00}\times10^{-5}\ M_\star$ 
for the inner 3:2 mode. KOI-227.01's mass is $\simeq 2.3\times10^{-4}\ M_\star$ for the outer 2:1 mode, 
$\simeq 5.9\times10^{-5}\ M_\star$ for the outer 3:2 mode, and $4.82_{-0.84}^{+1.00}\times10^{-5}\ M_\star$
for the inner 3:2 mode. With $M_\star=0.49\ M_\odot$ from Huber et al. (2014) (Table \ref{stellar}), this suggests 
mass $\simeq2$-47$\ M_\oplus$ for KOI-227.01, and $\simeq6$-36$\ M_\oplus$ for the companion (hereafter KOI-227.02). 
Since these masses, and also those indicated by other {\tt MultiNest} fits discussed above, are firmly in 
the planetary range, we believe that it is reasonably obvious that KOI-227 is a system of (at least) 
two planets. 

Table \ref{koi227_tab} lists the dynamical parameters of KOI-227.01 and 227.02 obtained from 
${\cal M}_{\rm V}$. With the physical radius of KOI-227.01 obtained from the transit analysis, 
$R_{01}\simeq2.2\, R_\oplus$, the dynamical fit implies physical density of $\rho_{01}\simeq19$ 
g cm$^{-3}$ for the outer 2:1 mode, $\rho_{01}\simeq5$ g cm$^{-3}$ for the outer 3:2 mode, and  
$\rho_{01}\simeq2$ g cm$^{-3}$ for the inner 3:2 mode. The former value would indicate a very dense transiter. 
Using models from Zeng \& Sasselov (2013), we estimate that the only way to explain the 2:1 mode is essentially 
a pure-iron planet. This would be inconsistent with the expectation that above $\sim1.75 \, R_\oplus$ 
planets tend to be more like mini-Neptunes than dense super-Earths (Lopez \& Fortney 2013). The outer 
3:2 mode would imply a predominantly rocky internal composition of KOI-227.01, while the inner 3:2 mode 
would require the presence of lighter elements and/or extended atmosphere (Zeng \& Sasselov 2013). 

The orbital period of KOI-227.02 at the reference epoch is found to be either 35.62, 26.58 or 11.83
days. The integration of orbits shows that the two planets are {\it inside} a resonance,
such that the resonant angle librates with a period of $\simeq4.5$-4.6~yr. This corresponds to the observed 
period of the dominant TTV term of KOI-227.01 (Figure \ref{koi227_fit}). An interesting property of the 
identified outer solutions is that KOI-227.01 should have an orbit with substantial eccentricity 
($e_{\rm min}\simeq0.3$ derived from the asterodensity profiling, see Table~\ref{params}, and $e_b \simeq0.2$-0.4 from 
the TTV analysis), potentially attained as a result of convergent migration of the two planets 
into a resonance.   

The orbital periods, semimajor axes and eccentricities undergo large oscillations due to the resonant 
interaction of the two planets. For example, the amplitude of the orbital period oscillations 
in the 2:1 resonance is $\simeq0.05$ day for KOI-227.01 and $\simeq0.35$ day for KOI-227.02.
We therefore point out that there is, in this case, a substantial difference between the period 
$P_c$ inferred from the dynamical fit at epoch $\tau_0=2454969.644331$~BJD$_{\rm UTC}$ (as reported in 
Table~\ref{koi227_tab}) and the mean period of KOI-227.02. Given the resonant configuration of the 
two planets, the mean period of KOI-227.02 would be $2\times17.68856=35.3772$ days for the outer 2:1 
resonance, $1.5\times17.68856=26.5329$ days for the outer 3:2 resonance, or $17.68856/1.5=11.7923$ days 
for the inner 3:2 resonance. More data will be needed to distinguish between these possibilities.
\section{Conclusions}
The main goal of this paper was to illustrate the TTV inversion method in diverse cases, where the
distribution of dynamical parameters obtained from the TTVs (and TDVs) is unimodal (KOI-884), bimodal (KOI-319), 
or multimodal (KOI-227). These cases document different challenges for the photo-dynamical analysis
of transits. The analysis of KOI-884 leads to a unique-period solution, mainly because the existing TTVs 
from Kepler have very good S/N ($\sim$55), good coverage (almost two super-periods of the near-resonant term), 
and resolve the short-periodic chopping produced by the orbital conjunctions between planets. This case is similar 
to KOI-142 (Nesvorn\'y et al. 2013, Barros et al. 2014), where a unique dynamical solution was obtained with 
a similar setup. 

KOI-319, on the other hand, shows TTVs with a small amplitude because the two planets are not
close to a major resonance. The TTV signal is irregular and apparently contains several short-periodic 
frequencies that are well resolved, because the corresponding periods are shorter than the 
observational baseline.  
This contributes to the success of the inversion method that was capable of ruling out most orbital
configurations, except for two planetary solutions flipped around the outer 2:1 resonance. This rather 
common degeneracy results from the fact that companions symmetrically placed around the 2:1 resonance 
will produce the same TTV frequencies (KOI-872 presented the same difficulty, except that the period 
degeneracy could have been resolved by using strict TDV constraints that were available in that case; 
Nesvorn\'y et al. 2012). 

\begin{figure}[t]
\epsscale{0.7}
\plotone{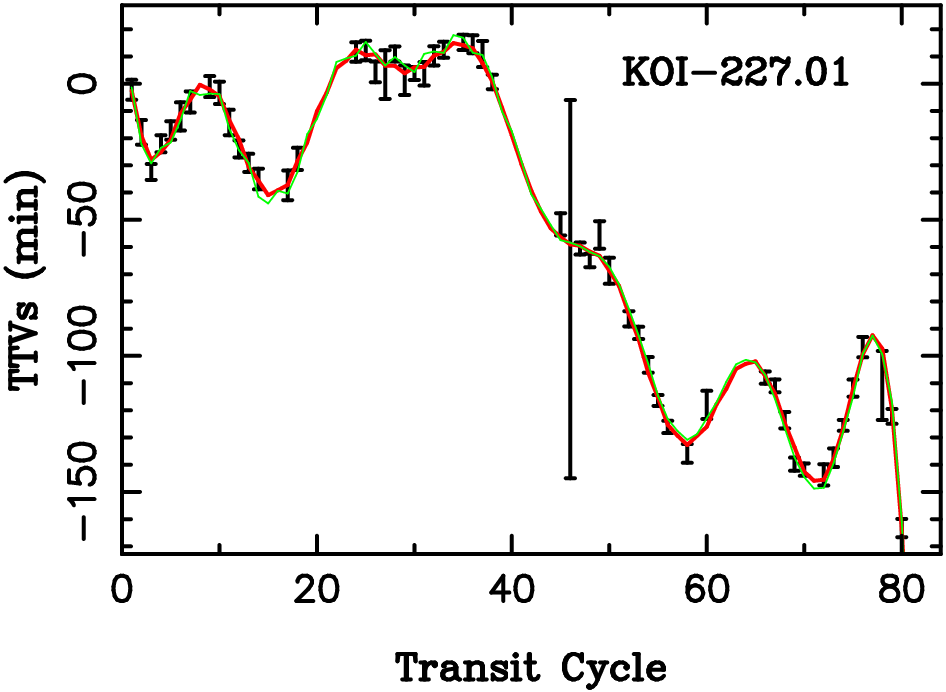}
\caption{TTVs of KOI-227.01. Here we removed the leading Fourier terms from the TTVs
to show the underlying signal. The red and green lines show the best fits corresponding to
the 2:1 and 3:2 resonances, respectively.}
\label{chop}
\end{figure}

The third system analyzed here, KOI-227, shows a huge TTV amplitude ($\simeq$10 hours), excellent
S/N ($\sim$150), and yet the TTVs inversion method fails to identify a unique solution. How is this 
possible? First, the time coverage of the TTV signal is still too short in this case as it does not
cover even one full period of the dominant TTV term. Second, the short-periodic chopping due to planet 
conjunctions is not resolved (Figure \ref{chop}). This happens probably because the outer companion 
has a relatively low mass such that the chopping is smaller than the measurement errors. The large TTV 
amplitude of KOI-227.01 is produced by resonant librations rather than by a large perturbing mass. 

Obtaining RV measurements would be desirable for all these systems. In the case of KOI-319, the RV
measurements are feasible because the host star is relatively bright (12.7 Kepler magnitude), and the 
expected RV amplitudes are $\sim$50 m s$^{-1}$ for KOI-319b, and 3.5 or 12 m s$^{-1}$ for KOI-319c. 
For KOI-227 and KOI-884, however, the RV measurements will be extremely challenging, if at all 
practical with the existing instrumentation, mainly because the host stars are faint (14.3 and 15.1 mag). 

As a final remark, we find it interesting how diverse the planetary systems discussed here are. In 
KOI-884, there is a Neptune- or Saturn-class planet on inside, as indicated by the physical 
radius of KOI-884b inferred from transits ($R \simeq 4.5\, R_\oplus$), and a massive 
$\simeq2.6\, M_{\rm J}$ planet on outside, with the two planets very near (but not in) the 3:1 resonance
($P_c/P_b=2.93$). In KOI-319, there is a Jupiter-class planet on inside ($R \simeq 10.7\, R_\oplus$),
and a Neptune- to Saturn-mass planet on a non-resonant, outer orbit. The two planets in KOI-227 are
probably super-Earths or Neptunes, and their orbits must be strictly resonant (most likely in the 
2:1 or 3:2 resonances). This diversity may be related to properties of the host stars (e.g., KOI-227 is 
an M-dwarf with $M_\star\simeq0.5\, M_\odot$), different conditions in the parent protoplanetary disks, 
or may signal a stochastic component of the planet formation and/or evolution. 
\acknowledgements
We thank the Kepler Science Team, especially the DAWG, for making the data used here available.
DMK is supported by the NASA Sagan fellowship.
%

\begin{table*}
\caption{Transit parameter estimates for KOI-227.01, KOI-319.01 and KOI-884.02.
${\rm BKJD}_{\rm UTC}={\rm BJD}_{\rm UTC}-2$,454,833. Term $(\rho_{\star,obs}/\rho_{\star,tru})$
denotes the mean stellar density derived from the composite transit light curve divided
by that derived by Huber et al. (2014). Any deviation of this parameter from unity implies 
that one or more of the idealized assumptions made in the determination of $\rho_{\star,obs}$ 
are invalid. This parameter has been used to derive the expectation for the minimum 
eccentricity $e_{\mathrm{min}}$ as described in Kipping (2013).} 
\begin{center} 
\begin{tabular}{c c c c} 
\hline
KOI & 227.01 & 319.01 & 884.02 \\ [0.5ex] 
\hline
$P$ [day] \dotfill & $17.688556_{-0.000016}^{+0.000018}$ & $46.151192_{-0.000026}^{+0.000025}$ & $20.483725_{-0.000019}^{+0.000016}$ \\ 
$\tau$ [BKJD$_{\mathrm{UTC}}$] \dotfill & $136.47793_{-0.00081}^{+0.00093}$ & $176.62522_{-0.00036}^{+0.00037}$ & $178.62753_{-0.00069}^{+0.00081}$ \\
$(R_P/R_{\star})$ \dotfill & $0.04320_{-0.00094}^{+0.00183}$ & $0.04739_{-0.00016}^{+0.00018}$ & $0.0517_{-0.0013}^{+0.0027}$ \\
$\rho_{\star,obs}$\,[g\,cm$^{-3}$] \dotfill & $2.78_{-0.98}^{+0.73}$ & $0.2095_{-0.0037}^{+0.0031}$ & $3.39_{-1.37}^{+0.90}$ \\
$b$ \dotfill & $0.58_{-0.11}^{+0.13}$ & $0.9083_{-0.0014}^{+0.0015}$ & $0.61_{-0.10}^{+0.14}$ \\ 
$q_1$ \dotfill & $0.5540$ & $0.4369$ & $0.4838$ \\ 
$q_2$ \dotfill & $0.3418$ & $0.3704$ & $0.4082$ \\ 
\hline
$R_P$\,[$R_{\oplus}$] \dotfill & $2.23_{-0.25}^{+0.27}$ & $10.67_{-0.39}^{+0.40}$ & $4.50_{-0.48}^{+0.55}$ \\
$u_1$ \dotfill & $0.5087$ & $0.4897$ & $0.5679$ \\ 
$u_2$ \dotfill & $0.2356$ & $0.1714$ & $0.1277$ \\
$S_{\mathrm{eff}}$\,[$S_{\oplus}$] \dotfill & $6.38_{-0.97}^{+2.16}$ & $60.4_{-3.5}^{+3.7}$ & $15.9_{-3.0}^{+6.8}$ \\
$(\rho_{\star,obs}/\rho_{\star,tru})$ \dotfill & $0.41_{-0.16}^{+0.25}$ & $1.016_{-0.072}^{+0.079}$ & $1.31_{-0.54}^{+0.55}$ \\
$e_{\mathrm{min}}$ \dotfill & $0.29_{-0.14}^{+0.15}$ &$0.017_{-0.012}^{+0.018}$ & $0.130_{-0.088}^{+0.114}$ \\ [1ex]
\hline\hline 
\end{tabular}
\label{params} 
\end{center}
\end{table*}

\begin{table*}
\caption{Transit times and durations of KOI-227.01, KOI-319.01 and KOI-884.02. 
An electronic version of the full table is available for download from ApJ.}
\begin{center} 
\begin{tabular}{c c c c c} 
\hline
cycle & $\tau_i$ & $E(\tau_i)$ & $\tilde{T}_i$ & $E(\tilde{T}_i)$ \\ 
      & [day]    &  [day]         &  [day]        & [day]       \\
\hline
\multicolumn{5}{c}{KOI-227.01} \\
0  &     54969.6443  &    0.0026&	0.1389 &    0.0066 \\
1  &     54987.2807  &    0.0031&       0.1056 &    0.0372 \\
2  &     55004.9129  &    0.0020&       0.1331 &    0.0050\\
3  &     55022.5596  &    0.0022&       0.1287 &    0.0053\\
4  &     55040.2023  &    0.0024&	0.1303 &    0.0059\\
5  &     55057.8474  &    0.0036&       0.1279 &    0.0103\\
6  &     55075.4968  &    0.0027&       0.1360 &    0.0068\\
8  &     55110.8097  &    0.0027&       0.1371 &    0.0065\\
9  &     55128.4722  &    0.0029&       0.1287 &    0.0074\\
10 &     55146.1344  &    0.0032&       0.1282 &    0.0085\\
etc. \\
\hline 
\hline
\end{tabular}
\label{td} 
\end{center}
\end{table*}

\begin{table*}
\caption{Stellar parameters from Huber et al. (2014).} 
\begin{center} 
\begin{tabular}{c c c c} 
\hline
KOI & 227.01 & 319.01 & 884.02 \\ [0.5ex] 
\hline
$T_{\rm eff}$\,[K] \dotfill & $3745_{-59}^{+51}$ & $5880_{-87}^{+87}$ & $5094_{-134}^{+168}$ \\
$\log g$ [g\,cm$^{-2}$] \, \dotfill &  $4.784_{-0.070}^{+0.070}$ & $3.927_{-0.014}^{+0.014}$ & $4.589_{-0.102}^{+0.023}$ \\
${\rm [Fe/H]}$\, \dotfill & $-0.02_{-0.10}^{+0.10}$ & $0.16_{-0.10}^{+0.10}$ & $0.21_{-0.29}^{+0.19}$ \\
$R_\star$\,[$R_{\odot}$] \dotfill & $0.47_{-0.05}^{+0.05}$ & $2.064_{-0.076}^{+0.076}$ & $0.791_{-0.053}^{+0.106}$ \\
$M_\star$\,[$M_{\odot}$] \dotfill & $0.49_{-0.06}^{+0.06}$ & $1.325_{-0.096}^{+0.096}$ & $0.884_{-0.089}^{+0.047}$ \\
$\rho_{\star}$\,[g\,cm$^{-3}$] \dotfill & $6.65_{-2.27}^{+2.27}$ & $0.206_{-0.015}^{+0.015}$ & $2.522_{-0.75}^{+0.32}$ \\ 
\hline\hline 
\end{tabular}
\label{stellar} 
\end{center}
\end{table*}

\clearpage
\begin{table}[t]
\caption{Dynamical parameters (and their associated errors) of KOI-884b and c inferred from 
${\cal M}_{\rm H}$. Mode 1 denotes the solution with the low $i_{\rm Mutual}$ value. Mode 2 
corresponds to the solution with high $i_{\rm Mutual}$. The reference epoch of orbital parameters 
given here is $\tau_0=2454970.0$~BJD$_{\rm UTC}$, just before the first observed transit of
KOI-884b.  The first column gives the primary parameter index that identifies each parameter 
in Figures \ref{koi884_lowi} and \ref{koi884_higi}.}
\begin{center}
\begin{tabular}{llrr}
\hline
    &                   & Mode 1 & Mode 2 \\
\hline
(1) & $M_b$ [$M_\star$]  & $1.7_{-1.2}^{+2.2}\times10^{-4}$ & $6.9_{-4.8}^{+9.1}\times10^{-4}$ \\
(2) & $M_c$ [$M_\star$]  & $2.62_{-0.23}^{+0.24}\times10^{-3}$ & $3.18_{-0.10}^{+0.11}\times10^{-3}$ \\
(3) & $\lambda_c$ [$^\circ$]  & $292.5_{-2.5}^{+2.7}$ & $269.6_{-1.0}^{+1.0}$    \\
(4) & $P_c$ [day]       & $59.947_{-0.034}^{+0.057}$  & $60.16_{-0.11}^{+0.20}$  \\
(5) & $e_b$             & $0.102_{-0.011}^{+0.012}$  & $0.0086_{-0.0024}^{+0.0033}$         \\   
(6) & $e_c$             & $0.0415_{-0.0068}^{+0.0068}$   & $0.0680_{-0.0089}^{+0.0081}$          \\
(7) & $\varpi_b$ [$^\circ$]  & $95.2_{-6.7}^{+5.2}$  & $139_{-33}^{+22}$                  \\  
(8) & $\varpi_c$ [$^\circ$]  & $58.3_{-8.2}^{+7.3}$ & $204.4_{-4.9}^{+3.9}$                 \\    
(9) & $b_c$             & $2.11_{-0.79}^{+1.18}$   & $16.3_{-1.2}^{+1.1}$           \\
(10) &$\Delta \Omega$ [$^\circ$] & $-9.5_{-5.9}^{+5.7}$ & $37.1_{-1.6}^{+1.8}$ \\
(11) & $P_b$ [day]              & $20.4737_{-0.0049}^{+0.0045}$ & $20.4646_{-0.0017}^{+0.0017}$     \\
(12) & $t_b$ [day]              & $0.6663_{-0.0013}^{+0.0013}$  & $0.6642_{-0.0015}^{+0.0014}$      \\        
\hline
   & $i_{\rm Mutual}$ [$^\circ$] & $0.75_{-0.51}^{+0.91}$ & $13.53_{-0.81}^{+0.73}$ \\ 
   & $M_b$ [$M_{\oplus}$] & $50_{-35}^{+65}$ & $200_{-140}^{+270}$    \\
   & $\rho_b$ [g cm$^{-3}$]  & $3.1_{-2.2}^{+4.0}$  & $12.2_{-4.2}^{+6.6}$  \\ 
   &  $M_c$ [$M_{\rm J}$] & $2.42_{-0.27}^{+0.29}$ & $2.94_{-0.24}^{+0.25}$ \\
\hline\hline
\label{koi884_tab}
\end{tabular}
\end{center}
\end{table}

\clearpage
\begin{table}[t]
\caption{Dynamical parameters (and their associated errors) of KOI-319b and c inferred from 
${\cal M}_{\rm H}$. Mode 1 denotes the solution just wide of the 5:3 resonance. Mode 2 
corresponds to the solution wide of the 7:3 resonance. The reference epoch of orbital parameters 
given here is $\tau_0=2455009.0$~BJD$_{\rm UTC}$, just before the first observed transit
of KOI-319b. The first column gives the primary parameter index that identifies each parameter 
in Figures \ref{koi319_mode1} and \ref{koi319_mode2}. The upper limits on $M_b$ were derived 
from the stability analysis.}
\begin{center}
\begin{tabular}{llrr}
\hline
    &                   & Mode 1 & Mode 2 \\
\hline
(1) & $M_b$ [$M_\star$]  & $<8\times10^{-3}$ & $<12\times10^{-3}$\\
(2) & $M_c$ [$M_\star$]  & $6.30_{-0.56}^{+0.70}\times10^{-5}$ & $2.37_{-0.23}^{+0.27}\times10^{-4}$\\
(3) & $\lambda_c$ [$^\circ$]  & $155.6_{-2.0}^{+2.0}$ & $16.5_{-4.0}^{+5.6}$    \\
(4) & $P_c$ [day]       & $80.099_{-0.056}^{+0.056}$  & $109.20_{-0.28}^{+0.25}$   \\
(5) & $e_b$             & $0.034_{-0.019}^{+0.024}$  & $0.0109_{-0.0073}^{+0.0111}$           \\   
(6) & $e_c$             & $0.032_{-0.015}^{+0.018}$  & $0.0300_{-0.0088}^{+0.0098}$          \\
(7) & $\varpi_b$ [$^\circ$]  & $55_{-32}^{+48}$  & $117_{-84}^{+49}$               \\  
(8) & $\varpi_c$ [$^\circ$]  & $3_{-38}^{+30}$ & $103_{-21}^{+20}$                       \\    
(9) & $b_c$             & $2.12_{-0.72}^{+0.93}$  & $7.9_{-2.6}^{+2.2}$           \\
(10) & $\Delta \Omega$ [$^\circ$] & $32.8_{-8.9}^{+11.0}$   & $11.5_{-3.1}^{+5.7}$          \\
(11) & $P_b$ [day]      & $46.14827_{-0.00085}^{+0.00071}$ & $46.1610_{-0.0040}^{+0.0034}$     \\
(12) & $t_b$ [day]      & $0.63114_{-0.00059}^{+0.00058}$  & $0.63171_{-0.00057}^{+0.00058}$      \\        
\hline
   & $i_{\rm Mutual}$ [$^\circ$] & $2.37_{-0.57}^{+0.91}$ & $7.3_{-2.7}^{+2.3}$ \\ 
   & $M_b$ [$M_{\rm J}$]  & $<10$ & $<16$    \\ 
   &  $M_c$ [$M_{\rm J}$]    & $27.7_{-3.4}^{+3.5}$ & $104_{-13}^{+14}$ \\
\hline\hline
\label{koi319_tab}
\end{tabular}
\end{center}
\end{table}

\clearpage
\begin{table}[t]
\caption{Dynamical parameters of KOI-227. The values and their uncertainties are shown for three parameter modes. 
The reference epoch of orbital parameters given here is $\tau_0=2454969.644331$~BJD$_{\rm UTC}$, corresponding to 
the first observed transit.}
\begin{center}
\begin{tabular}{llrrr}
\hline
    &                   &  outer 2:1 mode             & outer 3:2 mode   & inner 3:2 mode\\
\hline
(1) & $M_{01}$ [$M_\star$]  & $2.30_{-0.44}^{+0.60}\times10^{-4}$ & $5.9_{-3.0}^{+3.7}\times10^{-5}$& $2.4_{-1.4}^{+2.0}\times10^{-5}$ \\
(2) & $M_{02}$ [$M_\star$]  & $1.57_{-0.52}^{+0.70}\times10^{-4}$ & $7.7_{-1.6}^{+1.8}\times10^{-5}$& $4.82_{-0.84}^{+1.00}\times10^{-5}$\\
(3) & $\lambda_{02}$ [$^\circ$]  & $333_{-46}^{+33}$       & $158_{-51}^{+26}$                  & $271_{-23}^{+63}$      \\
(4) & $P_{02}$ [day]       & $35.618_{-0.048}^{+0.054}$     & $26.582_{-0.027}^{+0.042}$           & $11.8322_{-0.0064}^{+0.0093}$     \\
(5) & $e_{01}$             & $0.257_{-0.081}^{+0.099}$      & $0.35_{-0.21}^{+0.15}$               & $0.080_{-0.054}^{+0.077}$       \\   
(6) & $e_{02}$             & $0.030_{-0.014}^{+0.017}$      & $0.12_{-0.09}^{+0.14}$               & $0.28_{-0.15}^{+0.17}$   \\
(7) & $\varpi_{01}$ [$^\circ$]  & $346_{-79}^{+39}$        & $111_{-19}^{+59}$                    & $301_{-78}^{+85}$   \\  
(8) & $\varpi_{02}$ [$^\circ$]  & $285_{-57}^{+70}$        & $261_{-116}^{+54}$                   & $131_{-67}^{+84}$     \\    
(9) & $b_{02}$             & $16_{-10}^{+19}$           & $34_{-11}^{+14}$                       & $26_{-11}^{+12}$     \\
(10) & $\Delta \Omega$ [$^\circ$] & $180_{-159}^{+160}$   & $193_{-169}^{+153}$                 & $184_{-158}^{+151}$    \\
\hline
   & $i_{\rm Mutual}$ [$^\circ$] & $14.2_{-7.9}^{+15.9}$  &  $38_{-10}^{+14}$                    &  $41_{-18}^{+19}$   \\
   & $M_{01}$ [$M_{\oplus}$]  & $37.5_{-9.2}^{+10.2}$ & $9.6_{-2.4}^{+2.6}$                        & $3.9_{-2.3}^{+3.2}$    \\
   & $\rho_{01}$ [g cm$^{-3}$] &  $18.6_{-6.6}^{+10.1}$  & $4.8_{-1.7}^{+2.7}$                    & $2.0_{-1.1}^{+1.7}$ \\ 
   &  $M_{02}$ [$M_\oplus$]    & $25.6_{-10.1}^{+10.8}$ & $12.6_{-2.9}^{+3.2}$                    & $7.9_{-1.9}^{+2.1}$ \\
\hline\hline
\label{koi227_tab}
\end{tabular}
\end{center}
\end{table}

\clearpage

\clearpage

\clearpage

\clearpage

\clearpage

\clearpage

\clearpage

\clearpage


\end{document}